\documentstyle[epsfig,a4p,12pt,cite]{article}

\newcommand{\PNxxx}    
    
\def\eplemi{\mbox{$\mathrm{e^+ e^-}$}}

\def\Zz{\mbox{$\mathrm{Z}^0$}}

\def\Bs{\mbox{$\mathrm{B_s}$}}

\def\Bd{\mbox{$\mathrm{B_d}$}}
\def\Bu{\mbox{$\mathrm{B_u}$}}

\def\Kp{\mbox{$\mathrm{ K^+}$}}

\def\Km{\mbox{$\mathrm{ K^-}$}}

\def\Ds{\mbox{$\mathrm{ D_s}$}}

\def\Ups4s{\mbox{$\Upsilon(4S)$}}
\def\etal{\mbox{{\it et al.}}}

\def\GeVcc{\mbox{GeV/$c^2$}}
\def\GeVc{\mbox{GeV/$c$}}
\def\GeV{\mbox{GeV}}
\def\MeVc{\mbox{MeV/$c$}}
\def\MeVcc{\mbox{MeV/$c^2$}}

\newcommand{\inmath}[1] {\ifmmode#1\else$#1$\fi}
\newcommand{\definmath}[2] {\def#1{\ifmmode#2\else$#2$\fi}}
\definmath{\dEdx} {{\mathrm d}E/{\mathrm d}x}
\definmath{\PWpm} {\mathrm{W}^{\pm}}      
\definmath{\Pgtp} {\tau^{+}}        
\definmath{\Pgtm} {\tau^{-}}        
\definmath{\Pgtpm}   {\tau^{\pm}}         
\definmath{\Pgn}  {\nu}          
\definmath{\Pagn} {\overline{\nu}}     
\definmath{\Pq}      {\mathrm{q}}
\definmath{\Paq}  {\overline{\mathrm{q}}}
\definmath{\PQ}      {\mathrm{Q}}
\definmath{\PaQ}  {\overline{\mathrm{Q}}}
\definmath{\Pu}      {\mathrm{u}}
\definmath{\Pau}  {\overline{\mathrm{u}}}
\definmath{\Pd}      {\mathrm{d}}
\definmath{\Pad}  {\overline{\mathrm{d}}}
\definmath{\Ps}      {\mathrm{s}}
\definmath{\Pas}  {\overline{\mathrm{s}}}
\definmath{\Pc}      {\mathrm{c}}
\definmath{\Pac}  {\overline{\mathrm{c}}}
\definmath{\Pb}      {\mathrm{b}}
\definmath{\Pab}  {\overline{\mathrm{b}}}
\definmath{\Pt}      {\mathrm{t}}
\definmath{\Pat}  {\overline{\mathrm{t}}}
\definmath{\Pap}  {\overline{\mathrm{p}}}
\definmath{\Pan}  {\overline{\mathrm{n}}}
\definmath{\PaD}  {\overline{\mathrm{D}}}
\definmath{\PaDz} {\overline{\mathrm{D}}^{0}}
\definmath{\PaB}  {\overline{\mathrm{B}}}
\definmath{\PaBz} {\overline{\mathrm{B}}^{0}}
\definmath{\PsDpm}   {\mathrm{D}^{\pm}_{\mathrm{s}}}  
\definmath{\PcgLpm}  {\Lambda^{\pm}_{\mathrm{c}}}  
\definmath{\PDst} {\mathrm{D}^{*}}     
\definmath{\PKs} {\mathrm{K}^{0}_{\mathrm s}}     
\definmath{\PgLz} {\Lambda^{0}}        
\newcommand{\qqbar}  {\Pq\Paq}

\newcommand{\ccbar}  {\Pc\Pac}
\newcommand{\bbbar}  {\Pb\Pab}

\newcommand{\Ztobb}     {\Zz\to\bbbar}
\newcommand{\Ztocc}     {\Zz\to\ccbar}

\newcommand{\Gammaof}[1]   {\Gamma_{\!\smash{#1}\mathstrut}}

\newcommand{\Gcc}    {\Gammaof{\ccbar}}
\newcommand{\Gbb}    {\Gamma_{\mathrm b \overline{\mathrm b}}}
\newcommand{\Ghad}      {\Gamma_{\mathrm{had}}}
\newcommand{\GbbGhad}      {\Gbb/\Ghad}
\newcommand{\GccGhad}      {\Gcc/\Ghad}

\definmath{\GeV}  {\mathrm{GeV}}
\definmath{\GeVc} {{\mathrm{GeV}}\!/c}
\definmath{\GeVcc}   {{\mathrm{GeV}}\!/c^2}
\definmath{\MeV}  {\mathrm{MeV}}
\definmath{\MeVc} {{\mathrm{MeV}}\!/c}
\definmath{\MeVcc}   {{\mathrm{MeV}}\!/c^2}
\definmath{\MVm}  {\mathrm{MV}\!/\mathrm{m}}
\definmath{\keV}  {\mathrm{keV}}
\definmath{\keVcm}   {\mathrm{keV}\!/\mathrm{cm}}
\definmath{\kV}      {\mathrm{kV}}
\definmath{\km}      {\mathrm{km}}
\definmath{\meter}   {\mathrm{m}}
\definmath{\cm}      {\mathrm{cm}}
\definmath{\mm}      {\mathrm{mm}}
\definmath{\micron}  {\mu\mathrm{m}}
\definmath{\nm}      {\mathrm{nm}}
\definmath{\kg}      {\mathrm{kg}}
\definmath{\gram} {\mathrm{g}}
\definmath{\second}  {\mathrm{s}}
\definmath{\microsec}   {\mu\mathrm{s}}
\definmath{\degree}  {^\circ}
\definmath{\degC} {^\circ\mathrm{C}}
\definmath{\ohm}  {\Omega}
\definmath{\Mohm} {\mathrm{M}\Omega}
\definmath{\rad}  {\mathrm{rad}}
\definmath{\mrad} {\mathrm{mrad}}
\definmath{\nb}      {\mathrm{nb}}
\newcommand{\eqref}[1]  {(\ref{#1})}
\newcommand{\PhysLett}  {Phys.~Lett.}

\newcommand{\NIM} {Nucl.~Instr.\ Meth.}
\newcommand{\ZPhys}  {Z.~Phys.}

\newcommand{\OPALColl}  {OPAL Collab.}
\newcommand{\JADEColl}  {JADE Collab.}

\def\cent{\centerline}

\def\to{$\rightarrow$}
\def\Fi{$\phi$}
\def\fz{$\rm f_0(980)$}

\def\plm{$\pm$}

\begin{document}
\begin{titlepage}
\begin{center}
{\Large EUROPEAN ORGANIZATION FOR NUCLEAR RESEARCH}
\end{center}
\begin{flushright}
       CERN-EP/2000 - 065   \\  30 May 2000

\end{flushright}

\vspace{4cm}
 
\begin{center}
{\Huge\bf\boldmath First Measurement of the Inclusive\\ 
Branching Ratio of b Hadrons to\\ 
$\phi$ Mesons in Z$^{\mathbf 0}$ Decays}
\end{center}\bigskip\bigskip
\begin{center}{\LARGE The OPAL Collaboration
}\end{center}
\vspace{1cm}

\vspace{1.5cm}
\cent{\large\bf Abstract}
The inclusive production rate of $\phi$ mesons from
the decay of b hadrons produced in \Zz\ decays was
measured 
 to be Br(b\to$\phi$X) =
$0.0282\pm0.0013~({\rm stat.})\pm0.0019~({\rm syst.}),$ 
using data collected by the OPAL detector at LEP.
\vspace{3cm}

\cent {\large Submitted to Physics Letters B}

\end{titlepage} 

\begin{center}{\Large        The OPAL Collaboration
}\end{center}\bigskip
\begin{center}{

G.\thinspace Abbiendi$^{  2}$,
K.\thinspace Ackerstaff$^{  8}$,
C.\thinspace Ainsley$^{  5}$,
P.F.\thinspace Akesson$^{  3}$,
G.\thinspace Alexander$^{ 22}$,
J.\thinspace Allison$^{ 16}$,
K.J.\thinspace Anderson$^{  9}$,
S.\thinspace Arcelli$^{ 17}$,
S.\thinspace Asai$^{ 23}$,
S.F.\thinspace Ashby$^{  1}$
% ...
}\end{center}\bigskip
\bigskip
$^{  1}$School of Physics and Astronomy, University of Birmingham,
Birmingham B15 2TT, UK
\newline
$^{  2}$Dipartimento di Fisica dell' Universit\`a di Bologna and INFN,
I-40126 Bologna, Italy
\newline
$^{  3}$Physikalisches Institut, Universit\"at Bonn,
D-53115 Bonn, Germany
\newline
% ...
$^{  5}$Cavendish Laboratory, Cambridge CB3 0HE, UK
\newline
% ...
$^{  8}$CERN, European Organisation for Nuclear Research,
CH-1211 Geneva 23, Switzerland
\newline
$^{  9}$Enrico Fermi Institute and Department of Physics,
University of Chicago, Chicago IL 60637, USA
\newline
% ...
$^{ 16}$Department of Physics, Schuster Laboratory, The University,
Manchester M13 9PL, UK
\newline
$^{ 17}$Department of Physics, University of Maryland,
College Park, MD 20742, USA
\newline
% ...
$^{ 22}$Department of Physics and Astronomy, Tel Aviv University,
Tel Aviv 69978, Israel
\newline
$^{ 23}$International Centre for Elementary Particle Physics and
Department of Physics, University of Tokyo, Tokyo 113-0033, and
Kobe University, Kobe 657-8501, Japan
\newline
% ...
Budapest, Hungary.

\section{Introduction}
\label{Sec:Intro}
In \Zz\ decays, \Fi\ mesons are produced from three sources: direct
production in the fragmentation process, decay of charm hadrons in
\Ztocc\ events and
from the decay of bottom hadrons. 
According to the spectator model, \Bu\ and \Bd\ mesons decay to a \Fi\
meson via the cascade decay, b\to c\to\Fi.
Previous measurements of the branching ratio Br(\Bu/ \Bd\to \Fi X)
were made by the CLEO 
\cite{cleo} and ARGUS \cite{argus}
collaborations and were averaged by the Particle Data Group to give
the result
Br(\Bu/\Bd\to \Fi X) = 0.035\plm0.007\cite{pdg}.

 At LEP,
where all b hadron species are
produced, another mechanism exists where a \Bs\ meson also decays
directly into a \Fi\ meson. 
A measurement of the \Fi\ production rate in inclusive b hadrons, and in
particular the \Fi\
rates of the individual b
hadron species, can be used as an input to improve our knowledge of
the \Bs\ oscillation frequency.

In this paper, we describe a measurement of the
production rate of \Fi\ mesons from b hadrons. This includes both
direct and cascade b hadron decays. 
\Fi\ mesons were identified in their
charged kaon decay mode, and those not coming from \Ztobb\ events were 
suppressed by utilizing the relatively long lifetime of the b hadrons.

\section{Hadronic Event Selection and Simulation}
 \label{Sec:EvtSel}
The data used for this analysis were collected by the OPAL detector 
\cite{OPAL} from \eplemi\ collisons at LEP between
1990 and 1995 running at centre-of-mass energies in the vicinity of
the \Zz\ peak.
Hadronic \Zz\ decays were selected using the number of tracks
and the visible energy in each event as described in reference
\cite{evsel}. 
This selection yielded 4.41 million hadronic events. 

For optimisation of the selection of events and for some of the
studies of systematic uncertainties, we generated 6.5 million 5-flavour
hadronic \Zz\ decays (referred to as \qqbar\ Monte Carlo). 
Monte Carlo events were also used to determine the selection efficiency for
signal and background. For determination of signal
efficiency, we generated
80,000 \Ztobb\ events in which at least one of the b hadrons contained a
\Fi\ meson in its decay chain. For background efficiency, we
generated 1 million \Ztocc\ events.

All these samples were 
generated with the JETSET
7.4 Monte Carlo program \cite{jetset} with parameters tuned to the OPAL
data \cite{OPALtune}. The heavy quark
fragmentation was parameterised by the fragmentation
function of Peterson \etal\ \cite{peterson}; all samples were
processed with the full
OPAL detector simulation package~\cite{gopal} and subjected to the
same reconstruction and analysis algorithms as the data.

\section{Analysis Procedure}
\label{Sec:Analysis}

In each event, tracks and electromagnetic clusters that were not
associated with a track were combined into jets, using the
invariant mass algorithm with the E0 recombination scheme
and the parameter 
$y_{\mathrm {cut}}$ set to 0.04 \cite{jade}.
The primary vertex of the 
event was reconstructed using the 
tracks and the knowledge of the position and 
spread of the \eplemi\ collision point \cite{vertex}.

\begin{figure}[h]
   \begin{center}\hspace*{-.4cm}\mbox{
        \epsfig{file=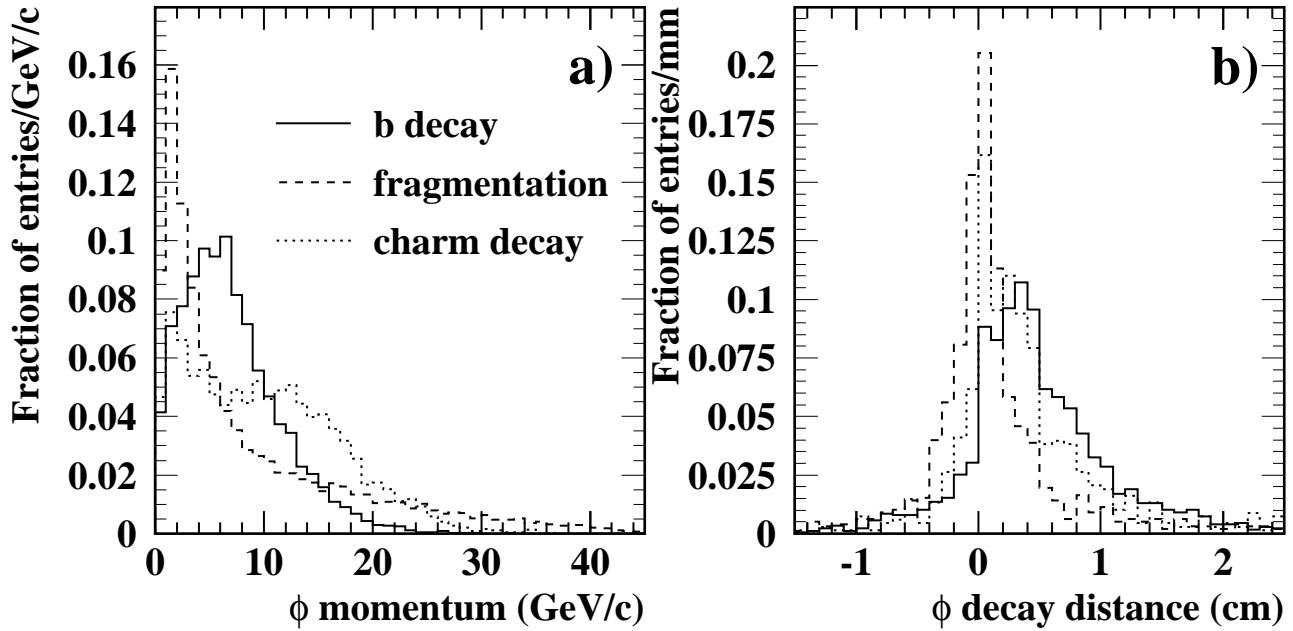}}
   \caption{Monte Carlo distributions of  (a) true momentum
     and (b) reconstructed decay distance of \Fi\ mesons from b decays
     (solid), fragmentation (dashed) 
     and charm decays (dotted).}
   \label{fig:treemom}
\end{center}
\end{figure}

We searched for the decay $\phi$\to\Kp\Km\
in the hadronic event sample 
by combining two tracks to form a $\phi$ meson
candidate. All two track combinations from the same jet with opposite
charges were considered.
Each of the two tracks was required to have a momentum of at least
2 \GeVc. Due to the small opening angle of
the two kaons, this momentum criterion is equivalent to a cut
on the \Fi\  momentum of about 4 \GeVc, which significantly reduces 
background from 
fragmentation while only accepting tracks in a momentum range with
good separation between kaons and pions using ionisation energy loss
(\dEdx) information \cite{dedx}.
The \dEdx\ for a kaon candidate track
was required to be consistent with that expected for a kaon with
a probability of more than 10\% as defined in \cite{dedx}.
To further reduce the pion background, the probability for a pion
hypothesis was required to be less than 10\%.
These \dEdx\ selection criteria retain 51\% of \Fi\ mesons in the
momentum range indicated above, while reducing the
combinatorial background by 97\%.
Figure \ref{fig:treemom} a) shows the momentum distribution of \Fi\ mesons
from fragmentation, b hadron and c hadron decays in
generator level simulation. 
The fragmentation \Fi\ meson
spectrum peaks at low momentum but also has a tail to high momentum
above the signal \Fi\ mesons. 
We therefore required the
momentum of the \Fi\ mesons to be smaller than 25 \GeVc.

Since at this point, the hadronic data sample consisted mostly of
non-\bbbar\ events, we suppressed these events by means of a b-tagging
algorithm, based on reconstructed displaced secondary vertices. 
An artificial neural network with inputs based on
decay length significance, vertex multiplicity and invariant mass 
information \cite{VNN}, was used to select
vertices with a high probability of coming from b hadron
decays. Events were accepted if at least one of the jets was tagged by the
neural network. The b-tagging selection was found to
be 68\% efficient for the signal, while rejecting 80\% of the 
remaining background.

The two kaon candidate tracks were fitted to a common vertex in three 
dimensions and the decay
distance, the distance from the primary vertex to the
reconstructed secondary vertex, was calculated.
Figure \ref{fig:treemom} b) compares the decay distance distribution of
\Fi\ mesons from fragmentation and from b and c decays which passed the above
selection criteria. In this plot, negative decay distance represents 
candidates with a reconstructed secondary vertex in the hemisphere 
opposite to the candidate's jet.
Motivated by this plot, we rejected
candidates with a negative decay distance. 
This selection left 62\% of the remaining
background events, while keeping 81\% of the signal events.

\begin{figure}[]
   \begin{center}\hspace*{-.4cm}\mbox{
        \epsfig{file=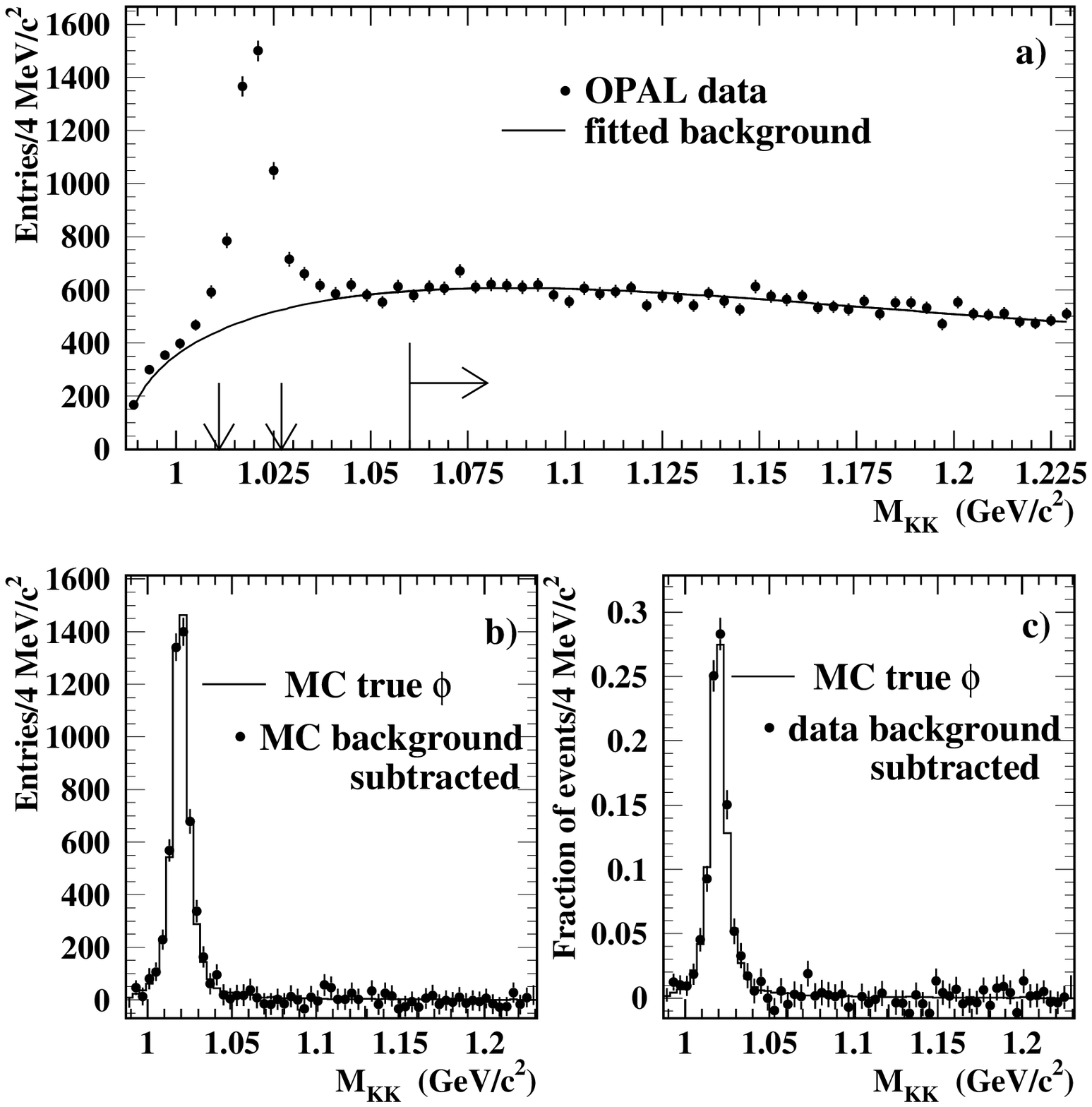}}
   \caption{a) Invariant mass distribution of candidates passing all the
     selection criteria. The solid line represents the combinatorial background
 estimation as described in the text. The arrows show the signal
 and fit normalisation regions. b) Comparison of the
 distribution of true Monte Carlo \Fi\
 mesons with that obtained by subtracting
 the combinatorial background shape from a fit to the combinatorial 
 background. c)
 Comparison of the
 distribution of true Monte Carlo
\Fi\ mesons with that obtained
 in the data after background-subtraction.
}
   \label{fig:phimass}
\end{center}
\end{figure}

Candidates were accepted if their invariant mass was in the region:  
$1.011~\GeVcc~<$ $~M_{\mathrm KK}~$ $<~1.027~\GeVcc,$ which
corresponds to about two standard deviations of the reconstructed
mass distribution around the 
nominal $\phi$ mass.
Figure \ref{fig:phimass} a) shows the invariant mass distribution of the
\Kp\Km\ candidates. A total of 4297 candidates were found in the
signal mass region.  

\subsection{Background Estimation}

Two of the three sources of $\phi$ mesons mentioned in section
\ref{Sec:Intro} are considered as background, in addition to the non-
$\phi$ background where the \Fi\ candidate is formed from a random
combination of tracks not both from a \Fi.
To estimate the combinatorial background, we fitted the $M_{\rm KK}$ 
distribution to the form
\begin{equation}
f(M_{\rm K}) = (M_{\rm KK}-2\cdot M_{\rm K})^a\cdot(b+c\cdot M_{\rm
  KK}+d\cdot M_{\rm KK}^2),
\end{equation}
where $M_{\rm K}$ is the mass of the charged kaon. The parameters
$a$, $b$, $c$, and $d$ were determined from a fit to the combinatorial
background in the \qqbar\ Monte Carlo sample, and then
the function shape was normalised to the data in the mass region
$1.060~\GeVcc~<$ $~M_{\mathrm KK}~$ $<~1.225~\GeVcc.$ 
The background shape is superimposed on Figure \ref{fig:phimass}
a). To verify that this technique estimates correctly  the background
shape, we compared the background-subtracted Monte Carlo sample with
the true distribution of Monte Carlo \Fi\ mesons (figure
\ref{fig:phimass} b). 
We also compared the
background-subtracted \Fi\ mesons in the data with the true Monte Carlo \Fi\
distribution (figure \ref{fig:phimass} c). Both sets of histograms agree with
each other very well, thus justifying the procedure for combinatorial
background estimation.
Subtracting the combinatorial background, we obtained $2667\pm40$ $\phi$ mesons
in the signal mass range, where the error arises from the statistical
uncertainty on the combinatorial background estimation.
The above background estimation ignores a possible small contribution
from \fz. As such a contribution is not well established, we
consider it as a systematic uncertainty as described in section \ref{Sec:Syst}.

The number above contains a small number of $\phi$ mesons from \Ztocc\
decays and from fragmentation. The former was estimated using known
and calculated branching ratios of ${\mathrm D_u,~D_d~ and~ D_s}$ into
$\phi$ mesons, as in \cite{blue}. This calculation is based on
isospin symmetry ratios between measured and theoretical branching
fractions and was updated to include recent results \cite{pdg}.
We used:
\begin{eqnarray}
{\rm Br(D_u\rightarrow\phi X)}& =&0.0182\pm 0.0025 , \\
{\rm Br(D_d\rightarrow\phi X)}& =&0.0182\pm 0.0025 , \\
{\rm Br(D_s\rightarrow\phi X)}& =&0.175\pm 0.025 . 
\end{eqnarray}
Using the partial widths of \Zz\ into \ccbar\ and $f_{\rm D_{u,d,s}}$, the branching
fractions of charm quark into the different D mesons as measured by LEP
\cite{LEPHLF}, we estimated the number of $\phi$ mesons which were
produced from charm quarks to be $52746\pm8322$, where the error
includes the uncertainty on all of the above numbers. With an overall
efficiency to survive the selection criteria estimated at 1.48\% and using
Br(\Fi\to\Kp\Km) = 0.491\plm0.008\cite{pdg}, we obtained an estimate
of $389\pm61$ $\phi$ mesons from this source. The error,
obtained from the above numbers, reflects
the systematic uncertainty on this estimation.

The fragmentation $\phi$ background was estimated directly from the
\qqbar\ Monte Carlo which had been tuned to reproduce the
\Fi\ meson yield in the data \cite{OPALtune}. It had also been shown
\cite{phiprod}, that the shape of the Monte Carlo distribution agrees
 with that of the data, in particular in the region below 35 \GeV.  
We estimate the number of \Fi\ mesons from this source to be
485\plm20, where the error is from the statistical uncertainty
on the Monte Carlo sample used for this estimation.

\section{Result and Systematic Uncertainties}

The number of \Fi\ mesons which originated from the decay of a b
hadron was calculated by subtracting the estimated background
contribution
from the number of \Fi\ mesons seen in the data. The result is
 $N_\phi = 1793\pm 50$, where the error is from the statistical
 uncertainty on the background level. 
To calculate the branching fraction, we used:
\begin{equation}
{\mathrm Br(b\rightarrow\phi X)} = {N_\phi \over 2\cdot \GbbGhad\cdot
 N_{\rm had}\cdot
 \epsilon\cdot  {\rm Br(\phi\rightarrow K^+K^-)}},
\end{equation}
where $N_{\rm had}$ is the number of hadronic events and $\epsilon$ is the
selection efficiency for \Fi\ mesons which originated from b
hadron decay.
With 4.41 million hadronic events, $\GbbGhad=0.21642\pm0.00073$
  \cite{LEPHLF} and $\epsilon = 0.068\pm0.001,$ we obtained
$${\rm Br(b\rightarrow\phi X)} = 0.0282\pm0.0013\pm0.0019,$$
where the first error is statistical and the second is a 6.6\%
systematic error, obtained as discussed in detail below.

\subsection{Systematic Uncertainties}
\label{Sec:Syst}
Systematic uncertainties arise from the limited accuracy with
which the branching ratios used in this analysis are
known, from the uncertainty in the simulation used to
determine the efficiency and from the background estimation. 
All uncertainties quoted below are relative, i.e. with respect to 
the measured value of Br(b\to\Fi X).
Table 1 summarises the systematic uncertainties detailed below.

\begin{table}[htb]
\begin{center}
\begin{tabular}{|l|c|} \hline
Systematic Source & $\delta({\rm Br(b\rightarrow\phi X)})/
{\rm Br(b\rightarrow\phi  X)}$ (\%) \\
\hline
Branching fractions & 3.6 \\
\dEdx & 3.1 \\
Fragmentation \Fi\ & 2.8\\
Heavy quark fragmentation modelling & 2.4\\
Combinatorial background & 1.6 \\
Monte Carlo statistics & 1.4 \\
b lifetime and decay multiplicity & 1.2\\
Detector modelling & 1.2\\
\hline
Total & 6.6 \\
\hline
\end{tabular}
\caption{Systematic uncertainties}
\label{tab:dedx}
\end{center}
\end{table}

\subsubsection*{Branching fractions}
The uncertainty on the known branching fractions of $\phi\rightarrow
{\rm K^+K^-},$ $\GbbGhad,$ $\GccGhad,$ $f_{\rm D_{u,d,s}}$ and 
${\rm D_{u,d,s}}\rightarrow
\phi$ resulted in a 3.6\%\  uncertainty on the branching ratio of
b\to\Fi X.
As \Bs\ mesons also decay directly to \Fi\ mesons, the efficiency for
detecting \Fi\ mesons from the decay of a \Bs\ meson is slightly
different from that of B$_{\rm u,d}$. Since only a small fraction of
\Bs\to\Fi X is expected to come from direct decays, we
increased the fraction of cascade decays in the Monte Carlo
from the current value of 85\% to 100\%, and recalculated the
efficiency. We also
varied the production rate of b hadrons according to the known values 
\cite{LEPHLF}. We obtained an additional uncertainty on the 
efficiency of 0.3\%.

\subsubsection*{\boldmath Modelling of \dEdx}
\label{Sec:dedx}
To estimate the uncertainty arising from the modelling of the \dEdx\
selection criteria, we compared the efficiency of the \dEdx\ cuts in 
Monte Carlo
events and in data. 
We reconstructed \Fi\ mesons without the \dEdx\ criteria 
in data and in Monte Carlo and compared the \Fi\ yield to that
obtained in
section \ref{Sec:Analysis}.
We obtained an efficiency associated with the \dEdx\ selection
criteria of 49.5\% in data and 51.0\% in Monte
Carlo. Hence, the relative uncertainty on
the signal efficiency associated
with the \dEdx\ cuts is estimated at 3.1\%.
A similar technique was used recently by OPAL \cite{brpvgcc}, and
consistent uncertainties obtained.

\subsubsection*{\boldmath Fragmentation $\phi$ mesons}
The largest dependence on the Monte Carlo in estimating the
background levels is related to fragmentation \Fi\ mesons.
Our selection criteria were designed to
reject fragmentation \Fi\ mesons, thereby limiting the possible
uncertainty associated with their simulation.  
Since we are estimating the fragmentation \Fi\ background directly
from the Monte Carlo, we would like to establish the level of accuracy
of the simulation of this source.
We took the small difference of \Fi\ meson yield
between the Monte Carlo and the measured value \cite{pdg},
and obtained a relative uncertainty of 2.7\%. 

A possible shape 
difference will affect the selection efficiency due to the momentum cuts. 
We checked the effect 
of varying the Lund symmetric fragmentation parameters $a$ (0.11-0.32 
\GeV$^{-2}$), and $b$ (0.48-0.58 \GeV$^{-2}$). These changes produced
a negligible effect. We also varied the parton shower $\Lambda$ value
in the range 0.13 to 0.31 \GeV. This resulted in a relative
uncertainty on the fragmentation \Fi\ production rate of 1.6\%. 
We also compared the
shape obtained when using the HERWIG event generator \cite{herwig},
and obtained a relative uncertainty of 2.4\%. 
The uncertainty on the production rate of \Fi\ mesons from the decay of
b hadrons associated with the estimation of the fragmentation
\Fi\ background was estimated at 2.8\%.

\begin{figure}[]
   \begin{center}\hspace*{-.4cm}\mbox{
        \epsfig{file=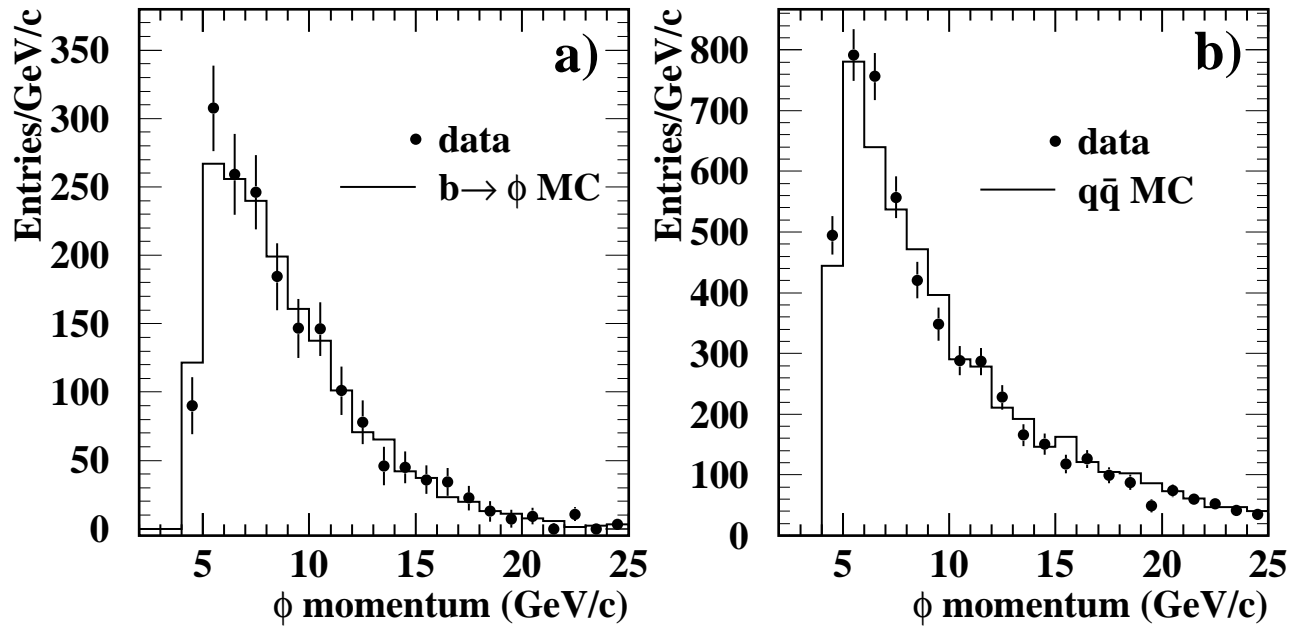}}
   \caption{Momentum distributions of (a) data \Fi\ mesons after
     background-subtraction (dots) and \bbbar\  Monte Carlo 
     (histogram), and (b) data and Monte Carlo candidates from 
     events that were
     rejected by the b-tagger or by the decay distance cut.
     }
   \label{fig:momcomp}
\end{center}
\end{figure}

To establish the agreement between the data and the Monte Carlo we
show in figure
 \ref{fig:momcomp} a) the momentum distribution of 
background-subtracted data and that of simulated \Fi\ mesons from the
decay of b hadrons normalised to the data area.  We also compared
(figure \ref{fig:momcomp} b) the momentum distribution of \Fi\ samples
which came mostly from fragmentation by selecting
events that were 
rejected by the b-tagger or by the decay distance cut. The data are
shown after combinatorial background-subtraction and the Monte Carlo
histograms are normalised to the same number of events passing the
hadronic event selection. According to the
Monte Carlo, 80\% of these \Fi\ mesons are from fragmentation.

\subsubsection*{Heavy quark fragmentation modelling}
Heavy quark fragmentation was simulated using the
function of Peterson~\etal~\cite{peterson}.
The heavy quark fragmentation model parameters
were varied to change the mean scaled
energy of weakly-decaying bottom and charm hadrons within the
experimental range: 
$\langle x_{E}\rangle_{\rm b}=0.702\pm 0.008$ and 
$\langle x_{E}\rangle_{\rm c}=0.484\pm 0.008$\cite{LEPHLF}. These
changes resulted in a 2.4\%  relative uncertainty.

\subsubsection*{Combinatorial background estimation uncertainty}
\label{Sec:sysbgfit}
The uncertainty on the fitted shape parameters and on the normalisation
gave an uncertainty on the background estimate.
We also repeated the procedure with a function of the form
$f(M_{\rm K}) = (M_{\rm KK}-2\cdot M_{\rm K})^a\cdot(b+c\cdot M_{\rm
  KK}),$ ($d=0$ in equation 1). This resulted
in 35 additional events or 1.5\%  relative uncertainty.
We also used
the Monte Carlo shape directly (without fitting), by normalising the
number of combinations in the mass range $1.06~<~ M_{\rm KK}~<~1.2~\GeVcc$
to that found in the data. With this method, we obtained 1681 events
(34 events more than our initial estimate).
Since the above two estimates are similar, we took the larger of the two
and combined it with the 0.5\%\ of uncertainty from the fitting
parameters, to obtain 1.6\%\ uncertainty on the branching ratio 
associated with the background estimation.

As the contribution from \fz\ was neglected in the background
estimation, we assigned a conservatively large estimation of this 
contribution as a 
systematic uncertainty.  We assumed that the width of
the \fz\ is 100 MeV and that Br(\fz\to K$\overline{\rm K}$) = 20\%. We used an
\fz\ multiplicity of 0.14 mesons per hadronic event \cite{pdg} and
obtained the selection efficiency from the Monte Carlo. Combining
these numbers, we estimate the contribution of \fz\ to the \Kp\Km\
invariant mass in the signal region to be 6 events. Assuming a 100\%
uncertainty in this estimate, this translates to
0.3\% of relative uncertainty on the branching fraction of b\to\Fi X.


\subsubsection*{b hadron lifetime and decay multiplicity}
The probability to reconstruct \Fi\  mesons from b hadron decays also
depends on the efficiency to reconstruct secondary
vertices in both hemispheres. This in turn is sensitive to the
charged decay multiplicity and lifetime of the b hadrons. The
Monte Carlo was reweighted to reflect the measured multiplicities and
lifetimes \cite{LEPHLF}. The uncertainty on these
figures gave an uncertainty of 1.2\% on the selection efficiency.

\subsubsection*{Detector modelling}
The simulated resolutions of the tracking detectors were varied by 
$\pm 10\thinspace\%$
relative to the values that optimally describe the
data, following the studies in~\cite{Rbmain}.
The analysis was repeated and the efficiencies were estimated again. 
This source contributed an uncertainty of 1.2\%.

\section{Summary and Discussion}
We have searched for direct and cascade decays of b hadrons into \Fi\ mesons.
The \Fi\ meson was reconstructed in its \Kp\Km\ decay mode and the
fraction of \Ztobb\ events in the sample was enriched. 
We found the branching fraction for
this process to be:  $${\rm Br(b}\rightarrow
\phi {\rm X}) = 0.0282\pm0.0013\pm0.0019.$$

Of special interest is the production rate of \Fi\ mesons from \Bs\
mesons. However, as the existing average of
B$^{\pm/0}\rightarrow \phi {\mathrm X}$ measurements has 20\%
uncertainty, it is impossible to obtain a meaningful result for  
Br(\Bs\to\Fi X).
On the other hand, our result may be used to obtain
an upper limit on the branching ratio of
B$^{\pm/0}\rightarrow \phi {\mathrm X}$. We considered the extreme 
scenario in which \Bs\ mesons
decay to \Fi\ mesons only via \Ds\ decays and used 
Br(\Bs\to\Ds X) = (90$\pm$33)\%\cite{pdg}. We made a conservative
assumption that b baryons do not contribute to the \Fi\ production rate 
and obtained an upper limit of:
Br(B$^{\pm/0}\rightarrow \phi {\rm X})\leq0.0285$ at the 90\%\ confidence
level. This limit is consistent with the CLEO result  
(0.023\plm0.006\plm0.005) \cite{cleo}, but
is only marginally consistent with that of ARGUS  (0.0390\plm0.0030\plm0.0035) \cite{argus}.

\section{Acknowledgements}

We particularly wish to thank the SL Division for the efficient operation
of the LEP accelerator at all energies
 and for their continuing close cooperation with
our experimental group.  We thank our colleagues from CEA, DAPNIA/SPP,
CE-Saclay for their efforts over the years on the time-of-flight and trigger
systems which we continue to use.  In addition to the support staff at our own
institutions we are pleased to acknowledge the  \\
Department of Energy, USA, \\
National Science Foundation, USA, \\
Particle Physics and Astronomy Research Council, UK, \\
Natural Sciences and Engineering Research Council, Canada, \\
Israel Science Foundation, administered by the Israel
Academy of Science and Humanities, \\
Minerva Gesellschaft, \\
Benoziyo Center for High Energy Physics,\\
Japanese Ministry of Education, Science and Culture (the
Monbusho) and a grant under the Monbusho International
Science Research Program,\\
Japanese Society for the Promotion of Science (JSPS),\\
German Israeli Bi-national Science Foundation (GIF), \\
Bundesministerium f\"ur Bildung und Forschung, Germany, \\
National Research Council of Canada, \\
Research Corporation, USA,\\
Hungarian Foundation for Scientific Research, OTKA T-029328, 
T023793 and OTKA F-023259.\\

\end{document}